%% file: main.tex
\documentclass{article}
\usepackage{spconf,amsmath,amssymb,graphicx,hyperref,booktabs,pifont}
\usepackage{microtype}
\usepackage{multirow} 
\usepackage{booktabs} 
\usepackage{graphicx}
\usepackage{enumitem}
\usepackage{hyperref}
\usepackage{booktabs}
\usepackage{makecell}
\usepackage{siunitx}
\usepackage{pifont}
\usepackage{graphicx}
\usepackage{subcaption}
\usepackage{makecell}
\usepackage{pgfplots}
\pgfplotsset{compat=1.18} % Use the latest compatibility mode
\usetikzlibrary{patterns} % For hatch patterns, if needed
\usepackage{tikz}
\usepackage{cite}

\let\oldthebibliography\thebibliography
\let\endoldthebibliography\endthebibliography
\renewenvironment{thebibliography}[1]{%
  \oldthebibliography{#1}%
  \setlength{\itemsep}{0pt}%        
  \interlinepenalty=10000%          
}{\endoldthebibliography}

\newcommand{\shortname}{UniverSR}

\title{UniverSR: Unified and Versatile Audio Super-Resolution\\via Vocoder-Free Flow Matching}
\name{
    Woongjib Choi, Sangmin Lee, Hyungseob Lim, Hong-Goo Kang
}
\address{
    Dept. of Electrical \& Electronic Engineering, Yonsei University, Seoul, South Korea\\
}

\begin{document}
\ninept
\maketitle
\input{0_abstract.tex}

\input{Figures/fig1}
\input{1.tex}
\input{2.tex}
\input{3.tex}

\input{4.tex}

\input{5.tex}

% -------------------------------------------------------------------------
% \vfill
\pagebreak
% \footnotesize
\bibliographystyle{IEEEbib}
\bibliography{refs}

\end{document}

%% file: 0_abstract.tex
\begin{abstract}
In this paper, we present a vocoder-free framework for audio super-resolution that employs a flow matching generative model to capture the conditional distribution of complex-valued spectral coefficients. 
Unlike conventional two-stage diffusion-based approaches that predict a mel-spectrogram and then rely on a pre-trained neural vocoder to synthesize waveforms, our method directly reconstructs waveforms via the inverse Short-Time Fourier Transform (iSTFT), thereby eliminating the dependence on a separate vocoder. This design not only simplifies end-to-end optimization but also overcomes a critical bottleneck of two-stage pipelines, where the final audio quality is fundamentally constrained by vocoder performance. 
Experiments show that our model consistently produces high-fidelity 48 kHz audio across diverse upsampling factors, achieving state-of-the-art performance on both speech and general audio datasets.
\end{abstract}

%% file: Figures/fig1.tex
\begin{figure*}[t]
    \centering
    
    \begin{subfigure}[b]{0.49\linewidth}
        \centering
        \includegraphics[clip, width=\linewidth]{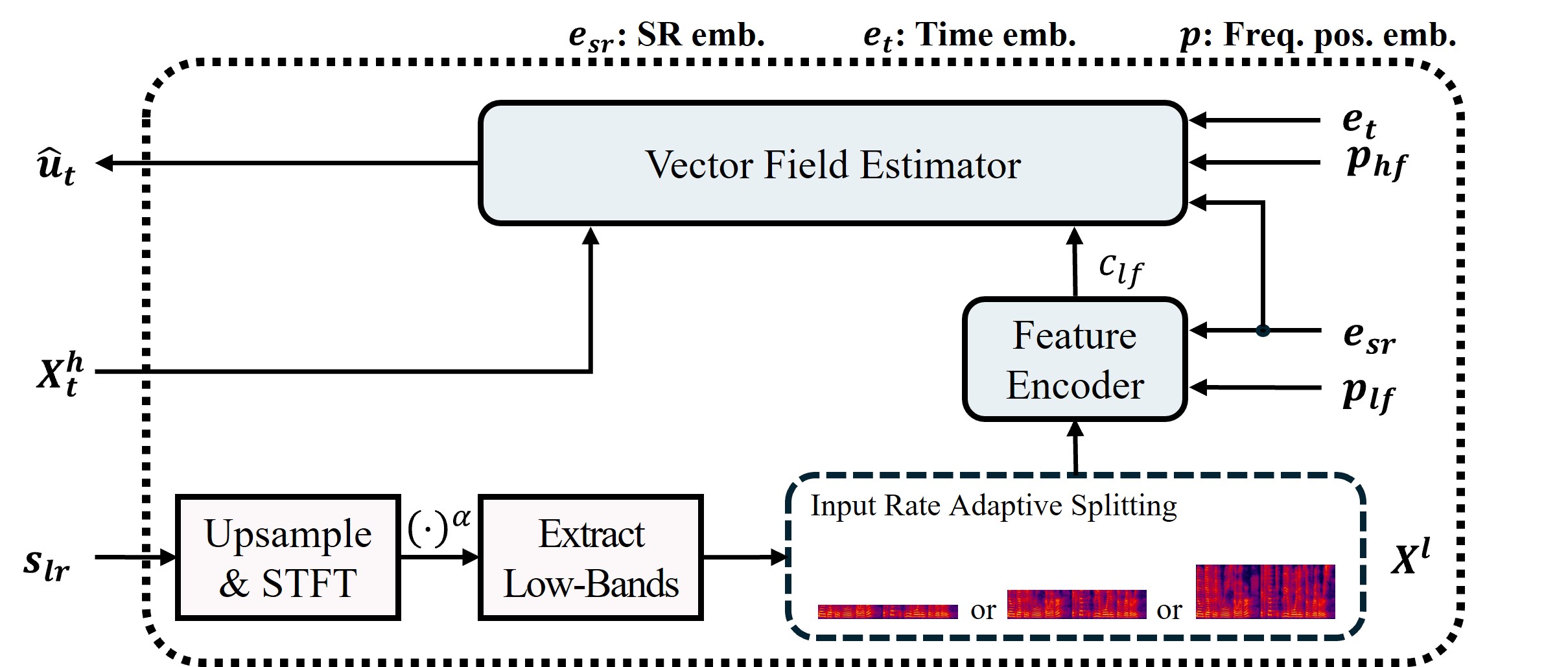} 
        \caption{Training stage}
        \label{fig:training_stage}
    \end{subfigure}
    \hfill % 
    \begin{subfigure}[b]{0.49\linewidth}
        \centering
        \includegraphics[clip, width=\linewidth]{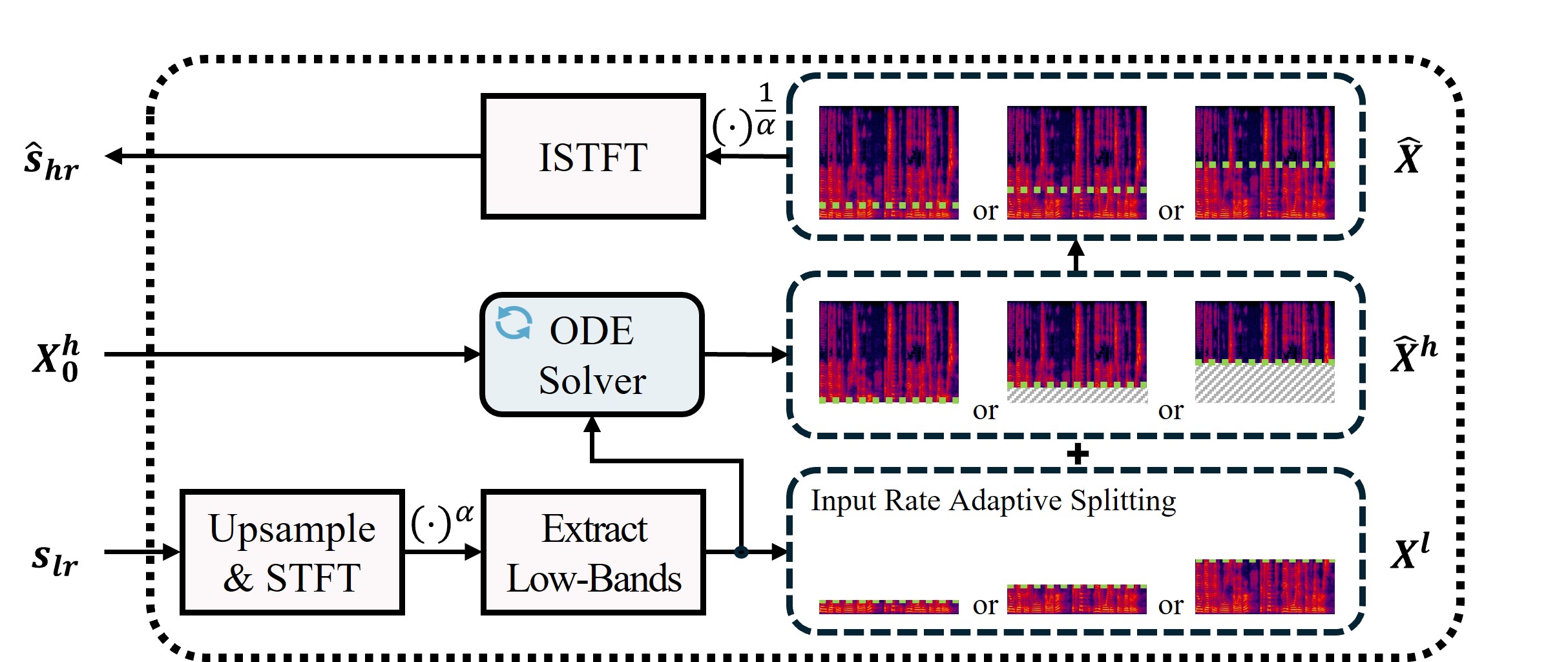}
        \caption{Inference stage}
        \label{fig:inference_stage}
    \end{subfigure}
    \vspace{-5pt}
    \caption{Overall framework of \shortname \, showing (a) training stage and (b) inference stage. Specifically, the ODE solver includes a feature encoder and vector field estimator.}
    \label{fig:fig}
    \vspace{-15pt}
\end{figure*}

%% file: 1.tex
\vspace{-5pt}
\section{Introduction}
\vspace{-3pt}
\label{sec:intro}
Increasing the sampling rate of audio signals has posed a fundamental challenge in signal processing, as communication channels, media streaming platforms, and storage devices impose strict bandwidth constraints.
When high-frequency components are absent, audio signals sound muffled and lack clarity. 
To address this issue, researchers have explored audio super-resolution (SR), also known as bandwidth extension (BWE), which reconstructs high-resolution (HR) audio from its low-resolution (LR) counterpart.
This is accomplished by estimating missing high-frequency content from band-limited representations through either signal processing techniques~\cite{larsen2005audio} or data-driven methods~\cite{kuleshov2017audiounet, lim2018tfnet}. 
Solving this problem supports applications such as enhancing speech intelligibility~\cite{li2019speechaudiosr, yu2024baenet} and restoring the fidelity of historical recordings~\cite{voicefixer, moliner2022behm}.

Recent advances in audio SR have been predominantly driven by generative models, which can be broadly categorized into one-stage (end-to-end) and two-stage pipelines. 
% Early approaches
Early end-to-end approaches \cite{kuleshov2017audiounet, lim2018tfnet} attempted to minimize an L2 reconstruction loss on the output waveform directly. However, these methods often produced over-smoothed results, lacking fine-grained textural details. 
% GAN
Subsequent approaches based on Generative Adversarial Networks (GANs) demonstrated substantial progress, with models such as Streaming SEANet \cite{li2021streamingseanet} operating directly on waveforms, while others, including AERO~\cite{mandel2023aero} and AP-BWE~\cite{lu2024apbwe}, focused on predicting spectral coefficients. 
% diffusion
Similarly, diffusion-based methods, including Schrödinger bridge variants~\cite{han2022nuwave2, yu2023udm+, li2025audio, kong2025a2sb}, have shown the ability to generate high-fidelity waveforms directly through multi-step sampling.
% limitation
However, one-stage generative approaches face distinct challenges: GANs suffer from training instability, often requiring carefully engineered losses and discriminators, while diffusion models are limited by severe inference inefficiency due to their iterative sampling process.

As an alternative to one-stage models, recent research has mostly opted for two-stage pipelines. 
Inspired by the success of mel-spectrogram-conditioned speech synthesis~\cite{kong2020hifi}, these methods decompose waveform reconstruction into two sub-tasks, in which an LR mel-spectrogram is first upsampled to its HR counterpart, and then a waveform is synthesized from the HR mel-spectrogram. 
Built on earlier approaches~\cite{voicefixer, liu2022nvsr}, 
\mbox{AudioSR}~\cite{liu2024audiosr} extended the two-stage paradigm to a latent diffusion-vocoder pipeline, enabling SR of general audio signals across diverse sampling rates.
Subsequent works~\cite{yun2025flowhigh, im2025flashsr} have further found improvements in reducing the number of sampling steps required for diffusion-based HR mel-spectrogram reconstruction. 
% transformer-based mel upsampling
More recently, Transformer-based architectures~\cite{frepainter, zhao2025hifisr} have been introduced to enable more robust extraction of intermediate features. 

% limitations of 2-stage approaches
However, the two-stage paradigm suffers from a fundamental bottleneck due to its reliance on mel-spectrograms as intermediate representations. Since phase information is omitted in mel-spectrograms, the quality of the final output depends heavily on the neural vocoder's ability to reconstruct a plausible phase~\cite{zhao2025hifisr, lee2025waveumamba}.
Furthermore, these approaches often require additional post-processing~\cite{liu2022nvsr, liu2024audiosr, yun2025flowhigh, frepainter}, such as replacing the low-frequency band of the generated signal with the original using the Short-Time Fourier Transform (STFT). 

In this paper, we propose \textbf{\shortname},\footnote{Demo: \url{https://woongzip1.github.io/universr-demo}}
\footnote{Code: \url{https://github.com/woongzip1/UniverSR}} a vocoder-free framework for \textbf{uni}fied and \textbf{ver}satile audio \textbf{s}uper-\textbf{r}esolution.
By utilizing flow matching~\cite{lipman2022flowmatching} in the spectral domain, our model directly estimates the conditional distribution of complex-valued spectral coefficients, enabling direct waveform reconstruction through the inverse STFT (iSTFT) without relying on a separate vocoder. 
The key contributions of this work are summarized as follows:
\vspace{-3pt}
\begin{itemize}[leftmargin=0.35cm] 
    \item  We propose a novel vocoder-free, end-to-end framework for audio SR that directly reconstructs waveforms without relying on a pre-trained neural vocoder.
\vspace{-2pt}
    \item By utilizing flow matching, our model achieves superior audio quality while requiring substantially fewer sampling steps compared to conventional diffusion-based approaches.
\vspace{-2pt}
    % add single stage
    \item Trained on a diverse audio dataset, our model achieves state-of-the-art quality for speech, music, and environmental sounds across multiple upsampling factors from $\times$2 to $\times$6.
\vspace{-2pt}
\end{itemize}

%% file: 2.tex
% \vspace{-5pt}
\section{Proposed Method}
\vspace{-3pt}
\label{sec:prop}
Fig.~\ref{fig:fig} illustrates the overall framework of UniverSR.
Given a low-resolution (LR) signal $s_{lr} \in \mathbb{R}^l$, the objective is to estimate the corresponding high-resolution (HR) version $s_{hr} \in \mathbb{R}^{l'}$, where $l$ and $l'$ are the number of samples in each waveform. 
% Upsample & STFT
The input $s_{lr}$ is first upsampled via sinc interpolation to match the target HR length $l'$.
This upsampled signal is then transformed into a complex spectrogram of shape $\mathbb{R}^{F \times T \times 2}$, where $F$ and $T$ denote the number of frequency bins and frames, respectively, and the last dimension represents the real and imaginary components. 
For notational simplicity, the batch dimension is omitted. 
To reduce dynamic variations across frequency bands, we apply a power-law compression $(\cdot)^\alpha$ to the magnitude of the spectrogram while preserving its original phase.
% STFT Extraction
We then extract the low-frequency bins that contain meaningful spectral content and obtain the low-band spectrum $X^l \in \mathbb{R}^{F_1 \times T \times 2}$.
Here, $F_1$ denotes the number of frequency bins up to the Nyquist frequency of the original LR input.

We frame audio super-resolution task as a spectrum inpainting problem~\cite{lim2018tfnet, frepainter}, where the goal is to predict the missing upper-band spectrum from the low-band spectrum $X^l$.
Since $F_1$ varies depending on the input signal's sampling rate, we define a fixed-size generation target $X^h \in \mathbb{R}^{(F-F_1^{min}) \times T \times 2}$, which covers all possible high-frequency regions.
The constant $F_1^{min}$ represents the number of frequency bins for the lowest input bandwidth (e.g., 4~kHz) supported by our model.
This generative process is achieved by training a vector field estimator (VFE) conditioned on $X^l$ using flow matching~\cite{lipman2022flowmatching}.
The final spectrum is reconstructed by concatenating $X^l$ with the last $(F-F_1)$ bins of $\hat{X}^h$, discarding the generated bins that overlap with the low-band spectrum.

\vspace{-5pt}
\subsection{Model Architecture}
\vspace{-3pt}
\label{sec:arch}
\noindent\textbf{Vector Field Estimator (VFE).}
As shown in Fig.~\ref{fig:architecture} (a), the VFE adopts a U-Net with 2D ConvNeXt V2 blocks~\cite{woo2023convnext} as a backbone to estimate the target vector field from the noisy high-frequency spectrogram $X_t^h$. 
The U-Net consists of an initial convolutional layer, a series of encoder blocks, a bottleneck block, and corresponding decoder blocks with skip connections.
Each encoder block is composed of several stacked ConvNeXt V2 blocks followed by a downsampling layer, which progressively halves the time-frequency resolution while doubling the number of feature channels. 
The decoder mirrors this structure with transposed convolutions to upsample the feature maps while reducing channel depth. 
The entire backbone is conditioned on a rich set of features, which are described next.
\input{Figures/fig2}

\vspace{2pt}
\noindent\textbf{Conditioning Mechanism.}
The VFE is conditioned on a rich set of features $\mathbf{c}$, including an acoustic representation from the low-band spectrum, frequency-positional embeddings, and global context embeddings for time and sampling rate.

The acoustic feature is a frame-wise representation $c_{lf} \in \mathbb{R}^{T \times D}$, where $D$ is the feature dimension.
As shown in Fig.~\ref{fig:architecture} (b), this feature is extracted from the low-band spectrogram $X^l$ using a dedicated \textit{feature encoder}.
To provide spectral location awareness, we employ a sinusoidal positional embedding $p\in \mathbb{R}^{F \times D}$~\cite{vaswani2017attention} for frequency bins. 
The feature encoder is conditioned on the low-frequency portion of this embedding, $p_{lf} \in \mathbb{R}^{F_1 \times D}$, along with a learnable sampling rate embedding $e_{sr}$, yielding a representation that incorporates both spectral position and input resolution.
The encoder employs adaptive pooling along the frequency axis to generate a fixed-dimensional output $c_{lf}$, independent of the input's frequency resolution.

The acoustic feature $c_{lf}$ and the high-frequency positional embedding $p_{hf} \in \mathbb{R}^{(F-F_1^{min}) \times D}$ are then used to condition the main input of the VFE. 
Specifically, $p_{hf}$ modulates the broadcasted $c_{lf}$ through Feature-wise Linear Modulation (FiLM)~\cite{perez2018film}, producing a spatial condition map with a shape of $\mathbb{R}^{(F-F_1^{min} )\times T \times D}$.
This spatial condition map is then concatenated with  $X_t^h$ along the channel axis to form the input to the U-Net. 
Finally, a global context embedding, obtained by summing the time embedding $e_t$ and the sampling rate embedding $e_{sr}$, is linearly projected and added to the feature maps within each ConvNeXt block of the U-Net backbone.

\vspace{-5pt}
\subsection{Flow Matching for Conditional Spectrum Generation}
\label{sec:fm}
\noindent\textbf{Conditional Probability Path.}
Following conditional flow matching (CFM)~\cite{lipman2022flowmatching}, 
let $x_1$ denote a sample from the target high-band spectrum $X^h$, 
and $x \sim \mathcal{N}(\mathbf{0}, I)$ a sample from the prior.
We first define a conditional probability path
$p_t(x|x_1) = \mathcal{N}(x; \mu_t x_1, \sigma_t^2 I)$.
A sample from this path can be obtained via the conditional flow $\psi_t$:
\begin{equation}
    \label{eq:flow}
    \psi_t(x) = \mu_t x_1 + \sigma_t x,
\end{equation}
where $\mu_t = t$ and $\sigma_t = 1-(1-\sigma_{\min})t$. 
Note that $X_t^h$ in Section~\ref{sec:arch} corresponds to $\psi_t(x)$.
The target vector field is given by:
\begin{equation}
    \label{eq:target_vf}
    u_t(x | x_1) = \frac{d\psi_t(x)}{dt} = x_1 - (1-\sigma_{\min})x.
\end{equation}

\noindent\textbf{Training Objective.}
The VFE $v_\theta$, parameterized by $\theta$, is trained to approximate the target vector field by minimizing:
\begin{equation}
    \label{eq:cfm_loss}
    \mathcal{L}_{\text{CFM}} = \mathbb{E}_{t, p(\mathbf{c}, x_1), p(x)} 
    \left[ \| v_\theta(\psi_t(x), t, \mathbf{c}) - u_t(x | x_1) \|^2 \right],
\end{equation}
where $t \sim \mathcal{U}[0,1]$ and $\mathbf{c}$ is the conditioning set described in Section~\ref{sec:arch}.
To enable classifier-free guidance (CFG)~\cite{ho2022classifier}, we stochastically replace $c_{lf}$ with a null embedding during training.

\vspace{2pt}
\noindent\textbf{Inference Procedure.} 
Starting from $x\sim \mathcal{N}(\mathbf{0},I)$ (denoted as $X_0^h$ in Fig.~\ref{fig:fig} (b)), we solve the ordinary differential equation (ODE):
\vspace{-3pt}
\begin{equation}
    \label{eq:inference_ode}
    \frac{dx_t}{dt} = v_\theta(x_t, t, \mathbf{c}), \quad x_0 = x
\end{equation}
using a numerical ODE solver from $t=0$ to $t=1$.
To apply CFG, we replace $v_\theta$ with the guided vector field~\cite{zheng2023guided}:
\begin{equation}
    \label{eq:cfg}
    \tilde{v}_\theta(x_t, t, \mathbf{c}) = (1-w) v_\theta(x_t, t, \mathbf{c}_\emptyset) + w \cdot v_\theta(x_t, t, \mathbf{c}),
\end{equation}
where $\mathbf{c}_\emptyset$ denotes the conditioning set with $c_{lf}$ replaced by the null embedding and $w$ is the guidance scale.
The final state $\hat{x}_1$, denoted as $\hat{X}^h$, is cropped and concatenated with the low-band spectrum $X^l$ to form the full-band spectrum $\hat{X}$.
Finally, $\hat{X}$ is converted into the HR waveform $\hat{s}_{hr}$ via inverse power-law scaling and iSTFT.

%% file: Figures/fig2.tex
\begin{figure}[t]
    \centering
    % (a) First image (top)
    \begin{subfigure}[b]{0.48\textwidth}
        \centering
        \includegraphics[width=\linewidth]{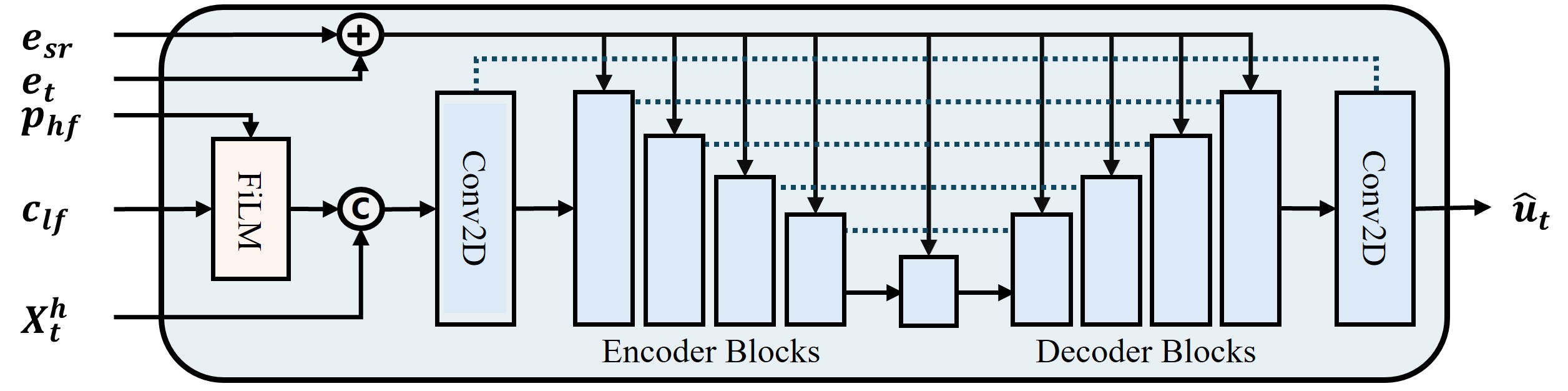} 
        \vspace{-0.5cm}
        \caption{Vector field estimator.}
        \label{fig:sub_a}
    \end{subfigure} 
    \vspace{0.1cm} 
    \vspace{-0.4cm}
    % (b) Second image (bottom)
    \begin{subfigure}[b]{0.48\textwidth}
        \centering
        \includegraphics[width=\linewidth]{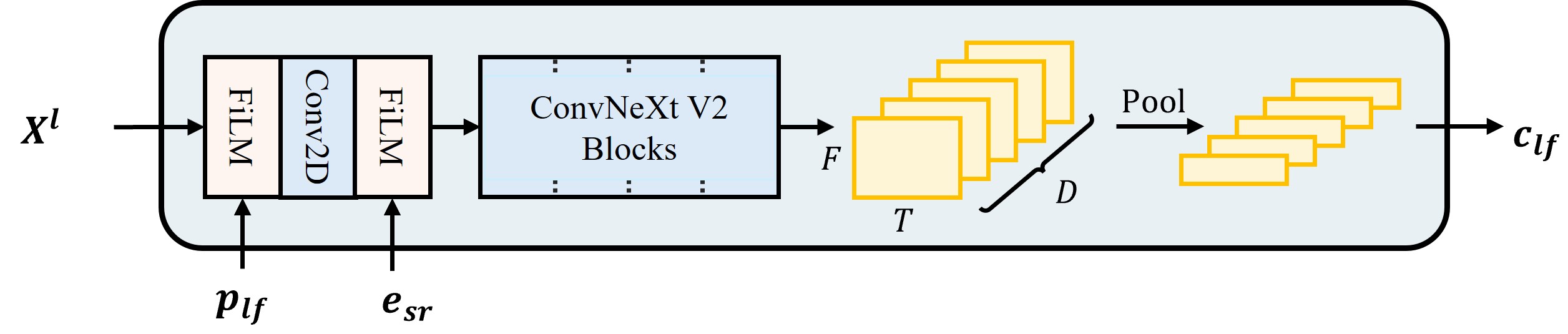} 
        \vspace{-0.9cm}
        \caption{Feature encoder.}
        \label{fig:sub_b}
    \end{subfigure}
    \vspace{-0.6cm} % Optional: adjust space before the main caption
    \caption{Detailed architecture of the (a) vector field estimator (VFE) and (b) feature encoder. Encoder, bottleneck, and decoder blocks of the VFE consist of a stack of ConvNeXt V2 blocks.
    }
    \label{fig:architecture}
    \vspace{-10pt} % Optional: adjust space after the figure
\end{figure}

%% file: 3.tex
\vspace{-5pt}
\section{Experiments}
\vspace{-3pt}

\input{Tables/table_audio}
\label{sec:intro}

\vspace{-7pt}
\subsection{Datasets}
\vspace{-3pt}
\label{sec:datasets}
We train two versions of our model to ensure fair and comprehensive evaluation. 
% Audio version
First, a single, unified model is trained on a diverse, aggregated corpus for robustness across multiple audio domains. The training data comprises three main categories: 1) \textbf{Speech} (218 hours from HQ-TTS~\cite{voicefixer}, EARS~\cite{ears}, and Expresso~\cite{expresso}); 2) \textbf{Music} (460~hours from Good-sounds~\cite{goodsounds}, MAESTRO~\cite{maestro}, MUSDB18~\cite{musdb}, \mbox{MedleyDB}~\cite{medleydb}, and MoisesDB~\cite{moisesdb}); and 3) \textbf{Sound Effects} (53 hours from FSD50K~\cite{fsd50k}). 
% VCTK version
Second, for a direct and fair comparison with existing speech-centric baseline models predominantly trained on VCTK~\cite{vctk}, we also train a specialized model exclusively on the VCTK training set.
For evaluation, we use a multi-domain test set consistent with the prior works~\cite{liu2024audiosr,im2025flashsr}. Specifically, we use 100 speech samples from VCTK~\cite{vctk}, a combined music set of 100 tracks from FMA-small~\cite{fma}, 100 instrumental pieces from URMP~\cite{urmp}, and 200 sound effects from ESC50 5-fold~\cite{esc50}. 
% For the VCTK-specialized model, we evaluate its performance on unseen speakers \texttt{p280} and \texttt{s5} from our VCTK test split.
The VCTK-specialized model is evaluated on speakers \texttt{p280} and \texttt{s5} from our VCTK test split, who were held out from the VCTK training set to assess generalization to unseen speakers.
\input{Figures/fig3_in_latex}
\input{Figures/fig4}

% Preprocessing
For preprocessing, all audio was resampled to 48\,kHz to serve as the HR ground truth. Segments with silence below -35 dB were then trimmed. LR inputs for training pairs were created by downsampling the HR signals after a low-pass filter based on a Hann window.
\vspace{-7pt}
\subsection{Implementation Details}
\vspace{-3pt}
Our model consists of a four-layer feature encoder with $D=384$ and a VFE with four encoder and decoder stages, which have ConvNeXt blocks with respective depths of [2, 2, 4, 2] and an initial channel size of 96.
This configuration yields a feature encoder with around 5M parameters and a VFE with around 52M parameters, totaling 57M parameters.
We use a 512-bin STFT representation with a window size of 1024 and 50\% overlap, where the last frequency bin is discarded. 
Additionally, we set a power compression ratio of $\alpha=0.2$ and $\sigma_{\min}=0.1$ for the CFM objective.

We train the model with AdamW optimizer with $\beta=(0.9, 0.999)$ and a learning rate of $2.0 \times 10^{-4}$ with a cosine decay schedule and 10k warmup steps. 
The unified and VCTK-specialized models are trained for 500k and 100k iterations, respectively. 
During training, the input sampling rate for each batch is randomly selected from {8, 12, 16, 24} kHz with probabilities of {0.7, 0.1, 0.1, 0.1}, corresponding to frequency cutoffs $F_1$ of {80, 128, 170, 256}, respectively. 
For the CFG, we use a conditioning dropout probability of 0.1 and a guidance scale $\omega$ of 1.5 for the four-step midpoint ODE solver during inference.

\vspace{-7pt}
\subsection{Evaluation Metrics}
\vspace{-3pt}
We adopt both objective and subjective metrics for our evaluations. 
% LSD
For objective assessment, we first measure Log Spectral Distance in the high-frequency bands (LSD-HF), a widely-used metric that calculates the distortion between the magnitude spectra of the target and generated audio in the upper frequency range.
% 2f
To better capture perceptual aspects, we also employ the 2f-model~\cite{2fmodel}, a pre-trained PEAQ-based estimator that estimates the mean MUSHRA score.
% MOS
Finally, for subjective validation, we conducted a listening test to gather Mean Opinion Score (MOS) ratings. In the test, 12 expert participants rated the perceptual audio quality on a scale from 1 to 5, evaluating 8 samples per model from each of the music, speech, and sound effect domains.

%% file: Tables/table_audio.tex
\newcommand{\cmark}{\ding{51}}%
\newcommand{\xmark}{\ding{55}}%
\begin{table}[t!]
\centering
\caption{
    Evaluation results for audio super-resolution models. 
    % L, 2f, and Voc. denote LSD-HF, 2f-model scores, and vocoder usage, respectively.
    L and 2f denote LSD-HF and 2f-model scores, respectively. 
}
\vspace{-5pt}
\label{tab:audio_evaluation_final}
\resizebox{\columnwidth}{!}{%
\begin{tabular}{
    c % Input rate
    l % Model
    r % Param. 
    c % Voc.
    % Speech (L, 2f)
    S[table-format=1.2] S[table-format=2.2]
    % Music (L, 2f)
    S[table-format=1.2] S[table-format=2.2]
    % Sound Effect (L, 2f)
    S[table-format=1.2] S[table-format=2.2]
}
\toprule
& & & & \multicolumn{2}{c}{\textbf{Speech}} & \multicolumn{2}{c}{\textbf{Music}} & \multicolumn{2}{c}{\textbf{Sound Effect}} \\
\cmidrule(lr){5-6} \cmidrule(lr){7-8} \cmidrule(lr){9-10}
\textbf{Input rate} & \textbf{Model} & \textbf{Size} & \textbf{Vocoder} & {L $\downarrow$} & {2f $\uparrow$} & {L $\downarrow$} & {2f $\uparrow$} & {L $\downarrow$} & {2f $\uparrow$} \\
\midrule
\multicolumn{2}{c}{\textbf{GT \small{(vocoded)}}} & -- & \cmark
& {0.67\textsuperscript{$\dagger$}} & {74.27} 
& {0.39\textsuperscript{$\dagger$}} & {69.32} 
& {0.46\textsuperscript{$\dagger$}} & {80.41} \\
\midrule
% 8kHz Results
\multirow{3}{*}{\textbf{8kHz}} 
& AudioSR~\cite{liu2024audiosr} & 672M & \cmark 
& 1.64 & \textbf{30.69} 
& 1.59 & 11.99 
& 1.52 & 22.58 \\
& FlashSR~\cite{im2025flashsr} & 639M & \cmark 
& 1.41 & 26.14 
& 1.31 & 18.01 
& 1.33 & 29.52 \\
& Proposed & 57M & \xmark 
& \textbf{1.40} & 26.58 
& \textbf{0.98} & \textbf{23.52} 
& \textbf{1.15} & \textbf{32.79} \\
\midrule
% 12kHz Results
\multirow{3}{*}{\textbf{12kHz}} 
& AudioSR~\cite{liu2024audiosr} & 672M &\cmark 
& 1.74 & 30.69 
& 1.51 & 14.22 
& 1.53 & 26.00 \\
& FlashSR~\cite{im2025flashsr} & 639M
& \cmark 
& 1.37 & 28.66 
& 1.41 & 20.46 
& 1.39 & 33.54 \\
& Proposed & 57M
& \xmark 
& \textbf{1.33} & \textbf{32.81} 
& \textbf{0.92} & \textbf{27.99} 
& \textbf{1.09} & \textbf{38.09} \\
\midrule
% 16kHz Results
\multirow{3}{*}{\textbf{16kHz}}
& AudioSR~\cite{liu2024audiosr} & 672M
& \cmark & 1.65 & 35.28 & 1.48 & 16.78 & 1.57 & 28.29 \\
& FlashSR~\cite{im2025flashsr} & 639M
& \cmark & \textbf{1.29} & 33.98 & 1.48 & 24.71 & 1.56 & 37.97 \\
& Proposed & 57M
& \xmark & 1.30 & \textbf{37.08} & \textbf{0.93} & \textbf{30.19} & \textbf{1.05} & \textbf{41.66} \\
\midrule
% 24kHz Results
\multirow{3}{*}{\textbf{24kHz}}
& AudioSR~\cite{liu2024audiosr}  & 672M 
& \cmark & 1.52 & \textbf{44.17} & 1.47 & 20.17 & 1.66 & 34.80 \\
& FlashSR~\cite{im2025flashsr} & 639M
& \cmark & \textbf{1.22} & 37.79 & 1.62 & 27.36 & 1.50 & 42.48 \\
& Proposed & 57M
& \xmark 
& 1.24 & 43.76 & \textbf{0.96} & \textbf{33.58} & \textbf{1.19} & \textbf{48.04} \\
\bottomrule 
\end{tabular}
}
\parbox{\columnwidth}{\scriptsize \textsuperscript{$\dagger$} As LSD-HF varies with the input rate, the value is for the 8~kHz condition. 
% Voc. denotes the use of a vocoder.
}
\vspace{-20pt} 
\end{table}

%% file: Figures/fig3_in_latex.tex
\definecolor{gtcolor}{RGB}{64,64,64}
\definecolor{gtvocoded}{RGB}{153,153,153}
\definecolor{audiosr}{RGB}{162, 213, 216}
\definecolor{flashsr}{RGB}{108, 178, 168}
\definecolor{propcolor}{RGB}{58, 141, 161}
\begin{figure}[t!]
\centering
\vspace{-0.1cm}
\begin{tikzpicture}
\begin{axis}[
    ybar,
    bar width=6pt,
    ymin=0, ymax=5.0,
    ylabel={Mean Opinion Score (MOS)},
    ytick={0,1,2,3,4,5},
    xtick={1,3,5,7},
    xticklabels={Speech, Music, Sound Effect, Average},
    xtick style={draw=none},
    xticklabel style={font=\normalsize},
    yticklabel style={font=\normalsize},
    legend style={
        at={(0.5,1.05)},
        anchor=south,
        legend columns=-1,
        font=\footnotesize,
        column sep=3pt,
        draw=none, % label box
    },
    legend image code/.code={
      \draw[#1, draw=none] (0cm,-0.08cm) rectangle (0.18cm,0.08cm);
    },
    ymajorgrids,
    grid style={dashed,line width=0.5pt,gray,opacity=0.5},
    width=\columnwidth,
    height=5cm,
    enlarge x limits=0.2,
]
% --- Vertical separators between classes ---
\draw [gray, dashed, thick] (axis cs:2,0) -- (axis cs:2,5);
\draw [gray, dashed, thick] (axis cs:4,0) -- (axis cs:4,5);
\draw [gray, dashed, thick] (axis cs:6,0) -- (axis cs:6,5);
% --- GT ---
\addplot+[
    ybar, draw=none, fill=gtcolor,
    error bars/.cd, y dir=both, y explicit,
    error bar style={black, line width=1pt}
] coordinates {(1,4.74) +- (0,0.11) (3,4.48) +- (0,0.17) (5,4.45) +- (0,0.17) (7,4.56) +- (0,0.09)};
\addlegendentry{GT}
% --- GT (Vocoded) ---
\addplot+[
    ybar, draw=none, fill=gtvocoded,
    error bars/.cd, y dir=both, y explicit,
    error bar style={black, line width=1pt}
] coordinates {(1,3.64) +- (0,0.22) (3,4.30) +- (0,0.17) (5,4.34) +- (0,0.18) (7,4.09) +- (0,0.12)};
\addlegendentry{GT (Vocoded)}
% --- AudioSR ---
\addplot+[
    ybar, draw=none, fill=audiosr,
    error bars/.cd, y dir=both, y explicit,
    error bar style={black, line width=1pt}
] coordinates {(1,3.50) +- (0,0.23) (3,3.38) +- (0,0.24) (5,3.33) +- (0,0.25) (7,3.40) +- (0,0.14)};
\addlegendentry{AudioSR}
% --- FlashSR ---
\addplot+[
    ybar, draw=none, fill=flashsr,
    error bars/.cd, y dir=both, y explicit,
    error bar style={black, line width=1pt}
] coordinates {(1,3.31) +- (0,0.22) (3,3.54) +- (0,0.21) (5,3.86) +- (0,0.22) (7,3.57) +- (0,0.13)};
\addlegendentry{FlashSR}
% --- Prop ---
\addplot+[
    ybar, draw=none, fill=propcolor,
    error bars/.cd, y dir=both, y explicit,
    error bar style={black, line width=1pt}
] coordinates {(1,4.34) +- (0,0.19) (3,4.44) +- (0,0.18) (5,4.42) +- (0,0.17) (7,4.40) +- (0,0.10)};
\addlegendentry{Proposed}
\end{axis}
\end{tikzpicture}
\vspace{-10pt}
\caption{
Subjective evaluation results (MOS) with 95\% confidence intervals for 8~kHz to 48~kHz upsampling. Dashed lines indicate separation between classes.}
\label{fig:mos_result}
\vspace{-15pt}
\end{figure}
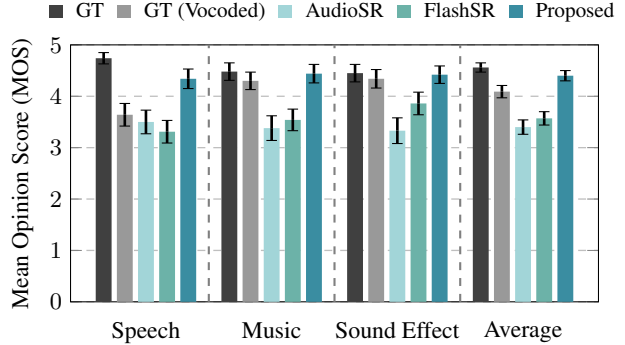

%% file: Figures/fig4.tex
\begin{figure*}[h!]
    \centering

    \begin{subfigure}[t]{0.139\textwidth}
        \includegraphics[width=\linewidth]{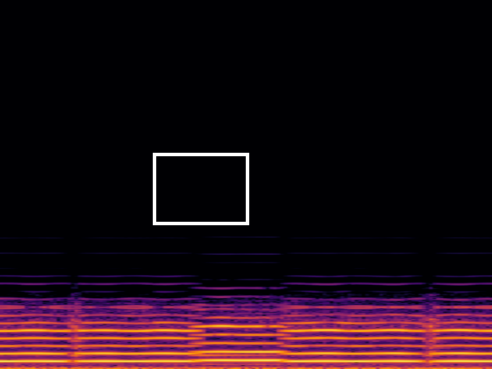}
    \end{subfigure}
    \hfill 
    \begin{subfigure}[t]{0.139\textwidth}
        \includegraphics[width=\linewidth]{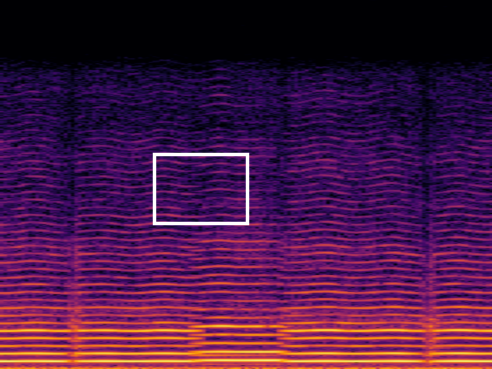}
    \end{subfigure}
    \hfill
    \begin{subfigure}[t]{0.139\textwidth}
        \includegraphics[width=\linewidth]{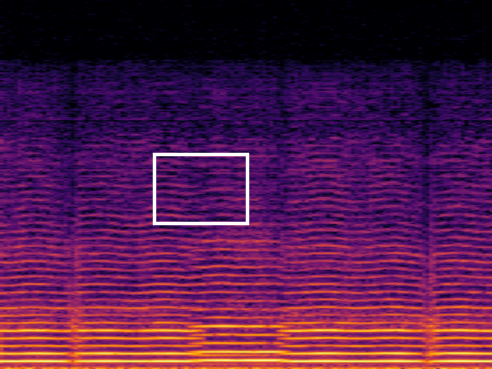}
    \end{subfigure}
    \hfill
    \begin{subfigure}[t]{0.139\textwidth}
        \includegraphics[width=\linewidth]{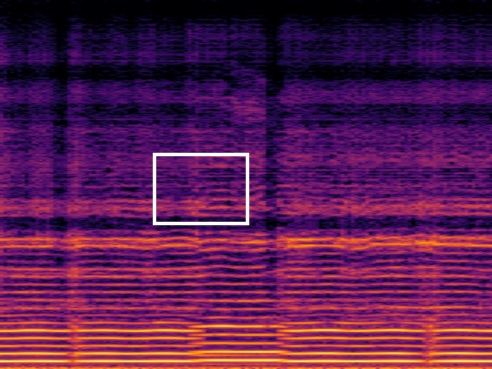}
    \end{subfigure}
    \hfill
    \begin{subfigure}[t]{0.139\textwidth}
        \includegraphics[width=\linewidth]{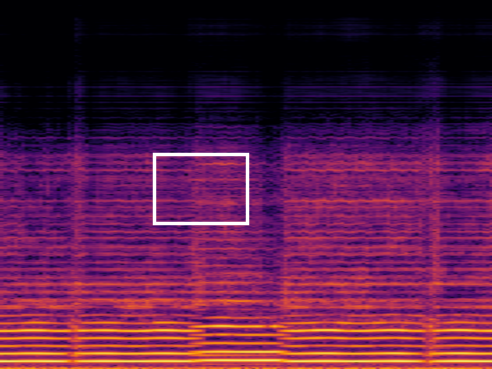}
    \end{subfigure}
    \hfill
    \begin{subfigure}[t]{0.139\textwidth}
        \includegraphics[width=\linewidth]{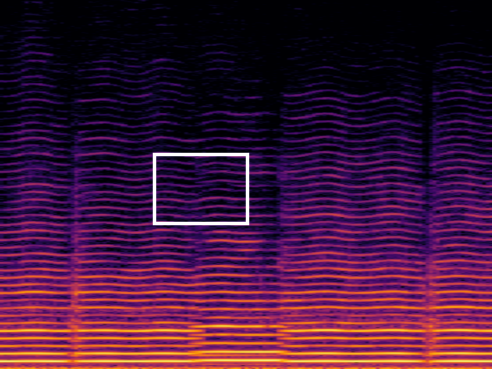}
    \end{subfigure}
    \hfill
    \begin{subfigure}[t]{0.139\textwidth}
        \includegraphics[width=\linewidth]{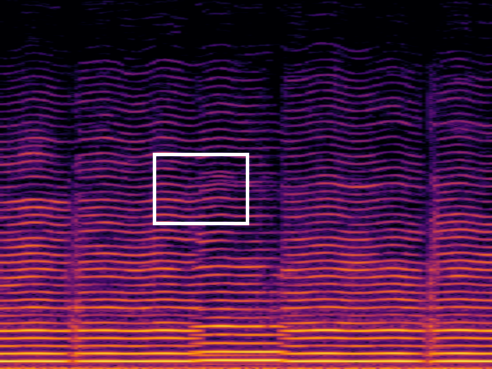}
    \end{subfigure}

    \vspace{1pt} % 

    \begin{subfigure}[t]{0.139\textwidth}
        \includegraphics[width=\linewidth]{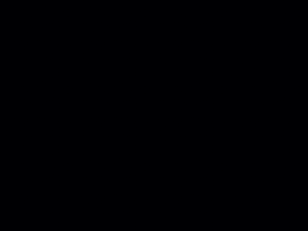}
        \caption{LR input}
        \label{fig:lr}
    \end{subfigure}
    \hfill 
    \begin{subfigure}[t]{0.139\textwidth}
        \includegraphics[width=\linewidth]{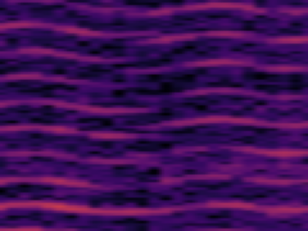}
        \caption{Ground Truth (GT)}
        \label{fig:gt}
    \end{subfigure}
    \hfill
    \begin{subfigure}[t]{0.139\textwidth}
        \includegraphics[width=\linewidth]{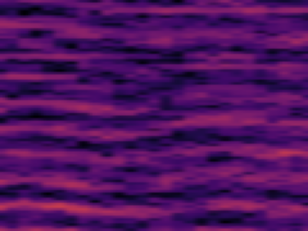}
        \caption{GT (Vocoded)}
        \label{fig:vocoded}
    \end{subfigure}
    \hfill
    \begin{subfigure}[t]{0.139\textwidth}
        \includegraphics[width=\linewidth]{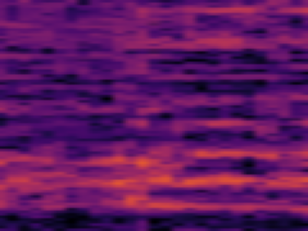} 
        \caption{AudioSR}
        \label{fig:audiosr}
    \end{subfigure}
    \hfill
    \begin{subfigure}[t]{0.139\textwidth}
        \includegraphics[width=\linewidth]{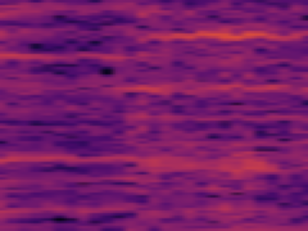} 
        \caption{FlashSR}
        \label{fig:flashsr}
    \end{subfigure}
    \hfill
    \begin{subfigure}[t]{0.139\textwidth}
        \includegraphics[width=\linewidth]{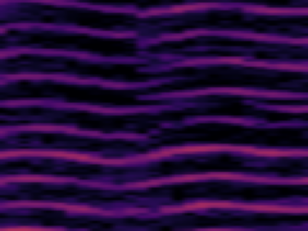} 
        \caption{\shortstack{Prop. ($\omega=1.5$)}}
        \label{fig:prop}
    \end{subfigure}
    \hfill
    \begin{subfigure}[t]{0.139\textwidth}
        \includegraphics[width=\linewidth]{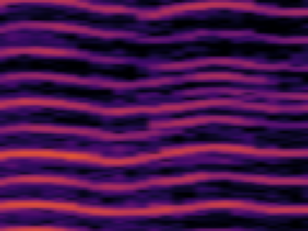} 
        \caption{\shortstack{Prop. ($\omega=2$)}}
        \label{fig:prop2}
    \end{subfigure}
    
    \vspace{-5pt}
    \caption{Spectrograms of a harmonic instrumental sample. 
     The bottom row displays magnified views of the regions enclosed by white rectangles in the top row.
     ``Prop." denotes our proposed model with a classifier-free guidance scale $\omega$.}
    \label{fig:spectrogram_comparison}
    \vspace{-15pt}
\end{figure*}

%% file: 4.tex
% \vspace{-5pt}
\section{Results and Analysis}
\vspace{-3pt}
\label{sec:results}
\subsection{Performance on Audio Super-Resolution}
\vspace{-3pt}
% Objective results
\noindent\textbf{Objective Evaluation.}
Table~\ref{tab:audio_evaluation_final} presents the objective evaluation results of our proposed model against vocoder-based audio super-resolution baselines: AudioSR~\cite{liu2024audiosr} and FlashSR~\cite{im2025flashsr}. 
To establish a practical upper bound regarding the reconstruction quality of these baseline models, we also include ground truth audio processed by the pre-trained vocoder from~\cite{liu2024audiosr} as `GT (vocoded)'. 
The results indicate that our model consistently outperforms the baselines in the music and sound effect domains across all sampling rates and metrics. 
For the speech domain, while our model demonstrates competitive LSD-HF scores, its 2f-model scores are slightly lower than the top-performing baseline under the 8~kHz and 24~kHz conditions.
% added
Notably, the baseline models require around 600M parameters due to their separate diffusion and vocoder components, whereas our unified architecture requires only 57M parameters.
\input{Tables/table_speech}

% MOS results
\vspace{2pt}
\noindent\textbf{Subjective Evaluation.}
To further assess perceptual quality, we conducted a subjective listening test (MOS) for the 8~kHz upsampling task. Results in Fig.~\ref{fig:mos_result} confirm that our proposed model achieves the highest average MOS score, indicating a clear preference by listeners. 
Particularly in the speech domain, despite its lower 2f-model score, our model's MOS score is not only significantly higher than the baselines but also surpasses that of the vocoded GT outputs.
We attribute this to the vocoder sometimes introducing subtle pitch instabilities when reconstructing harmonic-rich signals like speech, which can degrade the overall perceptual quality.

% Visualization results
\vspace{2pt}
\noindent\textbf{Qualitative Analysis.}
The higher performance of our model can be further illustrated by the spectrograms in Fig.~\ref{fig:spectrogram_comparison}. For a harmonic instrument, our proposed model demonstrates superior reconstruction of harmonic structures compared to the baselines. Notably, while the high-frequency components in the upper half of the vocoded GT are smeared and lack detail, our model generates cleaner and more structured high-frequency structures. This reveals an inherent limitation of vocoder-based approaches, in which their performance is upper-bounded by the capability of the vocoder they rely on.

\vspace{-10pt}
\subsection{Comparison with Speech Super-Resolution Baselines}
\vspace{-3pt}
For a direct comparison with speech-centric SR models, we trained our proposed model exclusively on the VCTK dataset. As shown in Table~\ref{tab_speech_in_domain}, we compare our model against vocoder-based (Fre-Painter~\cite{frepainter}, FlowHigh~\cite{yun2025flowhigh}) and single-stage diffusion (\mbox{NU-Wave2}~\cite{han2022nuwave2}, UDM+~\cite{yu2023udm+}) baselines, using ground truth samples reconstructed by FlowHigh's pre-trained vocoder as a practical upper bound for the vocoder-based approaches. 
While the vocoder-based models achieve competitive LSD-HF scores, they tend to produce overly smooth high-frequency components, resulting in lower perceptual quality scores compared to the diffusion-based approaches.
Meanwhile, our proposed model achieves the highest performance overall. 
Its superiority is particularly evident in the most challenging 8~kHz to 48~kHz upsampling task, where it achieves the best scores on both objective metrics. 
This result validates that our approach can achieve state-of-the-art speech restoration quality, even when trained on a domain-specific corpus.
\input{Tables/table_ablation}

\vspace{-10pt}
\subsection{Ablation Study}
\vspace{-3pt}
We conduct an ablation study to analyze the effect of the CFG scale, $\omega$. Our analysis reveals a trade-off between the perceptual richness of high-frequency components and the objective metric scores.
This improvement in perceptual quality is visually evident in the spectrograms in Fig.~\ref{fig:spectrogram_comparison} (f) and (g). The spectrogram generated with $\omega = 2.0$ clearly exhibits stronger and denser high-frequency structures compared to the one with $\omega = 1.5$. However, despite this perceptual richness, the objective metrics in Table~\ref{tab:audio_evaluation_final} are lower for $\omega=2.0$. This is because the generated signal deviates more from the ground-truth reference. Conversely, a scale of $\omega=1.0$ yields high objective scores but produces audibly flatter high-frequency components.
Therefore, selecting the $\omega$ scale involves balancing a trade-off between high-frequency expressiveness and source fidelity. While we use $\omega=1.5$ as a balanced default in this paper, this value can be tuned depending on the target audio domain and the user's specific goals.

%% file: Tables/table_speech.tex
\begin{table}[t!]
\vspace{3pt}
\caption{Evaluation results for speech super-resolution models. L and 2f denote LSD-HF and 2f-model scores, respectively. All models are open-sourced and trained with VCTK dataset. Best scores are in bold, second-best are underlined.}
\label{tab_speech_in_domain}
\vspace{-8pt}
\resizebox{\columnwidth}{!}
{
    \setlength{\tabcolsep}{4pt}
    \begin{tabular}{l c cc cc cc cc}
        \toprule
        % --- Table Header ---
        & 
        & \multicolumn{2}{c}{\textbf{8 $\rightarrow$ 48 kHz}} 
        & \multicolumn{2}{c}{\textbf{12 $\rightarrow$ 48 kHz}} 
        & \multicolumn{2}{c}{\textbf{16 $\rightarrow$ 48 kHz}} 
        & \multicolumn{2}{c}{\textbf{24 $\rightarrow$ 48 kHz}} \\
        \cmidrule(lr){3-4} \cmidrule(lr){5-6} \cmidrule(lr){7-8} \cmidrule(lr){9-10}
        \textbf{Model} & \textbf{Vocoder} & {L ↓} & {2f ↑} & {L ↓} & {2f ↑} & {L ↓} & {2f ↑} & {L ↓} & {2f ↑} \\
        \midrule

        % --- Table Body ---
        GT \small{(vocoded)} & \cmark
        & 0.66 & 79.05 & 
        0.67 & 79.05 & 
        0.68 & 79.05 & 
        0.70 & 79.05 \\
        \midrule

        Fre-Painter~\cite{frepainter} & \cmark
        & 1.25 & 27.02 & 1.23 & 29.50 & 1.18 & 31.43 & 1.07 & 35.16 \\

        FlowHigh~\cite{yun2025flowhigh} & \cmark
        & \underline{1.19} & 27.88 & \underline{1.17} & 30.66 & \underline{1.14} & 32.31 & 1.10 & 35.26 \\

        NU-Wave2~\cite{han2022nuwave2} & \xmark
        & 1.58 & 27.58 & 1.32 & 32.25 & 1.21 & 35.32 & 1.09 & 39.98 \\
        
        UDM+~\cite{yu2023udm+} & \xmark
        & 1.29 & \underline{29.12} & \textbf{1.16} & \underline{34.11} & \textbf{1.09} & \textbf{37.75} & \textbf{1.00} & \textbf{44.85} \\
        \midrule

        Proposed & \xmark
        & \textbf{1.14} & \textbf{31.41} & 1.20 & \textbf{34.42} & 1.17 & \underline{37.17} & \underline{1.06} & \underline{44.14} \\
        
        \bottomrule
    \end{tabular}
}
\vspace{-15pt}
\end{table}

%% file: Tables/table_ablation.tex
\begin{table}[t!]
\centering
\vspace{3pt}
\caption{
    Ablation study on the classifier-free-guidance (CFG) scale for 8~kHz to 48~kHz upsampling. Bold indicates the best performance. \,
    L and 2f denote LSD-HF and 2f-model scores, respectively.
}
\label{tab:ablation_cfg_scale}
\vspace{-7pt}
\resizebox{\columnwidth}{!}{% Adjust the resizebox width if needed
\begin{tabular}{
    c % CFG Scale
    % Speech
    S[table-format=1.2] S[table-format=2.2]
    % Music
    S[table-format=1.2] S[table-format=2.2]
    % Sound Effect
    S[table-format=1.2] S[table-format=2.2]
    % Average
    S[table-format=1.2] S[table-format=2.2]
}
\toprule
& \multicolumn{2}{c}{\textbf{Speech}} 
& \multicolumn{2}{c}{\textbf{Music}} 
& \multicolumn{2}{c}{\textbf{Sound Effect}} 
& \multicolumn{2}{c}{\textbf{Average}} % Average header added
\\
\cmidrule(lr){2-3} \cmidrule(lr){4-5} \cmidrule(lr){6-7} \cmidrule(lr){8-9} % c-midrule for Average added
\textbf{CFG Scale} & {L ↓} & {2f ↑} & {L ↓} & {2f ↑} & {L ↓} & {2f ↑} & {L ↓} & {2f ↑} \\ % sub-header for Average added
\midrule
% 8kHz Results
$\omega = 1.0$
% 1.0 
& 1.42 & \textbf{29.41} % Speech
& \textbf{0.92} & \textbf{25.22} % Music
& 1.16 & 32.65 % Sound Effect
& \textbf{1.07} & \textbf{28.24} \\ % Average values added
$\omega = 1.5$
% 1.5 
& \textbf{1.40} & 26.58 % Speech
& 0.98 & 23.52 % Music
& \textbf{1.15} & \textbf{32.79} % Sound Effect
& 1.10 & 26.95 \\ % Average values added
$\omega = 2.0$
% 2.0 
& 1.53 & 21.99 % Speech
& 1.09 & 21.32 % Music
& 1.21 & 31.46 % Sound Effect
& 1.20 & 24.65 \\ % Average values added
\bottomrule 
\end{tabular}
}
\vspace{-15pt}
\end{table}

%% file: 5.tex
\vspace{-3pt}
\section{Conclusion}
\vspace{-3pt}
\label{sec:conclusion}
In this paper, we introduced \textbf{\shortname}, a novel vocoder-free framework for audio super-resolution. Our model employs flow matching to learn the conditional distribution of complex-valued spectral coefficients, enabling direct waveform reconstruction through the inverse STFT. Trained on a large and diverse collection of audio datasets, our framework exhibits robust generalization performance across multiple domains and upsampling factors. Extensive objective and subjective evaluations demonstrate that UniverSR achieves state-of-the-art performance in upsampling 8, 12, 16, and 24 kHz audio to 48 kHz across speech, music, and environmental sound datasets.